\documentclass[showpacs,preprintnumbers,amsmath,amssymb]{revtex4}

\usepackage{graphicx}
\usepackage{dcolumn}
\usepackage{bm}

\begin{document}

\baselineskip=20pt
\vsize=490pt

\title{The Possibility of Existence of an Energy Gap in the One-Particle Excitation Spectrum of He II}
\author{I.M. Yurin}
\email{yurinoffice@mail.ru}
\affiliation{I.M.Yurin, Fl.61, bld. 7, 22 Festivalnaya St, Moscow, 125581, Russia}%
\author{S.A.Trigger}%
 \email{satron@mail.ru}
\affiliation{Joint Institute for High Temperatures, Russian Academy
of Sciences, 13/19, Izhorskaia Str.,
Moscow 127412, Russia}%

\begin{abstract}
Possibility for coexistence of a phonon and one-particle spectra
in ${}^4He$ at all temperatures including temperatures below the
$\lambda$ - point is shown. An approach developed to consider the
problem of the weakly imperfect Bose gas results in the existence
of a gap in the one-particle (one-atom) excitation spectrum below
the transition point. Thus, the microscopic model of He II turns
out to differ both from the model initially suggested by Landau
and from the Bogolyubov model accepted at present. Analysis of the
experimental data obtained by measuring enthalpy and the part of
helium atoms condensed in the state with the momentum $p=0$
enabled to estimate the size of the gap. It turned out to be
approximately equal to 0.31 K at zero temperature.

\end{abstract}

\maketitle
\section{\label{sec:level1}Introduction}

It is generally accepted that the superfluidity (SF) of He II was
explained by to L.D.Landau in 1941. Less known is the fact that
Landau initially assumed the existence of two branches in the
excitation spectrum of the system: phonon and roton branches. At
that, the latter corresponded to quantums of rotational degree of
freedom of the liquid and was separated from the ground state by an
energy gap.

 However, subsequently a viewpoint based, in particular, on the work of Bogolyubov \cite{1}
 was accepted. It assumes that only one phonon - roton excitation spectrum exists at zero temperature.
 We think that it is more correct to refer to it as one-particle (one-atom),
 because it is believed that the creation and annihilation of phonons
 cannot lead to a change in the number of atoms of the system,
 while the creation and annihilation operators for the so-called
 bogolons do not commute with the total number of particles in the system.
In particular, as a result, the basic Hamiltonian of the system
obtained by the Bogolyubov transformation does not preserve the
total number of atoms. This fact, however, is ignored by all the
authors: it is considered to be just a consequence of the
so-called spontaneous breaking of symmetry of the system due to
the presence of a second order phase transition.

This variant of describing the energy spectrum creates a number of
questions. On one hand, naturally, no one doubts the coexistence of
phonon and one-atom spectra at temperatures above the transition
point \cite{2}. So, there should be a mechanism of disappearance of
one of the excitation branches below the transition point. According
to the conventional point of view the disappearance of the phonon
branch occurs immediately after the appearance of a macroscopic
occupation number of the lower level of the system  \cite{2}. What
experimentalists consider to be a spectrum of phonons, strictly
speaking, as we mentioned above, is already a one-atom excitation
spectrum, although it is characterized by a linear relationship
between the excitation frequency and the wave vector in the
long-wave limit. Therefore, a gigantic hysteresis of the transition
temperature should exist in the system, because it is obvious that
the transition point in a boson system with one-atom spectrum
corresponding to the experimental phonon - roton spectrum of He II
should considerably exceed the value corresponding to the parabolic
spectrum of the system accepted for temperatures above the
transition point. In this connection it is pertinent to remind that,
according to the Landau theory, excitations of the phonon - roton
branch form a normal component in the two-fluid model of He II.

In discussing this problem, the systematic research on numerical
modelling of the properties of He II in the temperature range
where the system heat capacity considerably differs from the
phonon - roton gas heat capacity would be extremely interesting
and desirable. However, unfortunately, we do not know works in
this direction.

Quite recently the questions, concerning the concept itself of
spontaneous breaking of symmetry in the considering context, also
appeared \cite{3}. The point is that the total number of atoms in
the system is not preserved only in case of the truncated
Hamiltonian, in which only the expressions bilinear with respect to
the creation and annihilation operators of bogolons are selected. As
for the total Hamiltonian of the system, of course, it does not
break any symmetry. From this point of view the symmetry breaking is
rather mathematical in nature and associated with the approximations
inherent in the method used for calculating the wave functions of
the ground state and low-energy excitations. But on the physical
level, there is no symmetry breaking. If one does not take this into
account, a paradox \cite{3} results, which until now has not been
considered in the works concerning the description of the weakly
imperfect Bose gas as a qualitative model for the consideration of
physical processes in He II in the standard SF theory (of course, we
assume that the BCS - Bogolyubov theory of superconductivity is
obviously associated with the Landau - Bogolyubov SF theory).

Two possible options for further research are possible. On one
hand, one can try to prove that solving this paradox does not
affect the final results of calculations of He II within the
framework of the microscopic Bogolyubov theory. This method seems
to be very attractive, because a large number of experimental data
were described in the context of the standard approach (at least
qualitatively).

On the other hand, we think that it is very interesting to focus
efforts on another method: a search of options for describing the
system, which would maintain strictly the fundamental symmetries
of the system at all steps of considering it. It is in the search
for such an approach that we undertook this study. It will be
shown below that it suggests a different scheme of the excitations
branches in He II.

\section{\label{sec:level1}The Alternative Perturbation Theory in the Model of Weakly Imperfect Bose gas}
First, note that the result presented in this section was obtained
by Bogolyubov \cite{4}, if to limit to its mathematical part. The
limitation concerning the mathematical part is not accidental. It
is caused by the fact that, when Bogolyubov obtained a gap in the
spectrum of quasiparticles, he rejected this solution on the basis
of his concept of the physical properties of the system. We will
discuss this item later.

So, let us consider spinless bosons  (${}^4He$ atoms) with
interatomic interaction  $V\left( r \right)$ as a system located
in the box $L_x \times L_y  \times L_z$. We will assume that the
boundary conditions of the system are periodical. Then the
Hamiltonian of the system $\hat H$ in the model of the the weakly
imperfect Bose gas \cite{4} has the form:
\begin{equation}
\begin{gathered}
  \hat H = \hat T + \hat V, \hfill \\
  \hat T = \sum\limits_\mathbf{p} {t_\mathbf{p} b_\mathbf{p}^ +  b_\mathbf{p} } , \hfill \\
  \hat V = \frac{1}
{2}\Omega ^{ - 1} \sum\limits_{\mathbf{p},\mathbf{k},\mathbf{q}} {V_\mathbf{q} b_{\mathbf{p} + \mathbf{q}}^ +  b_{\mathbf{k} - \mathbf{q}}^ +  b_\mathbf{k} b_\mathbf{p} } , \hfill \\
\end{gathered}
\end{equation}
where $t_\mathbf{p}  = \frac{{p^2 }} {{2M}}$, $M$ is the atomic
mass of $^4He$, $\Omega$ is the system volume, $V_{\bf q}  = \int
{V\left( r \right)\cos \left( {{\bf qr}} \right)d{\bf r}}$ ,
$b_{\bf p}^+$ and $b_{\bf p}$ are the creation and the
annihilation operators for $^4He$ atoms in a state with a momentum
$\mathbf{p}$.

Let us calculate the one-atom energy $\tilde t_\mathbf{p}$ of the
$^4He$ atoms using an equation based on a random phase
approximation (RPA) $\left[ {\hat H,b_{\bf p}^ + } \right] =
\tilde t_{\bf p} b_{\bf p}^ +$. Let us select first the diagonal
part of the Hamiltonian $\hat H_{diag}$ , because only this part
is important for this calculation. For $\hat H_{diag}$ we easily
obtain:
\begin{equation}
\hat H_{diag}  = \hat T + \frac{1} {2}\Omega ^{ - 1}
\sum\limits_{\mathbf{p},\mathbf{k}} {\frac{{V_0  + V_{\mathbf{k} -
\mathbf{p}} }} {{1 + \delta _\mathbf{p}^\mathbf{k} }}b_\mathbf{p}^ +
b_\mathbf{k}^ +  b_\mathbf{k} b_\mathbf{p} } , \label{eq:two}
\end{equation}
where $\delta _{\bf p}^{\bf k}$ is the three-dimensional Kronecker
symbol. Denominator $1 + \delta _{\bf p}^{\bf k}$  was introduced in
 (\ref{eq:two}) in order to avoid the double account of the terms with
${\bf q} = 0$ at ${\bf p} = {\bf k}$.

Taking into account that, obviously,  $V_{\bf q}  = V_{ - {\bf
q}}$, equation $\left[ {\hat H_{diag} ,b_{\bf p}^ +  } \right] =
\tilde t_{\bf p} b_{\bf p}^+$ leads to the following expression
for $\tilde t_{\bf p}$:
\begin{equation}
\tilde t_{\bf p}  = t_{\bf p}  + \Omega ^{ - 1} \sum\limits_{\bf k}
{\frac{{V_0  + V_{{\bf k} - {\bf p}} }} {{1 + \delta _{\bf p}^{\bf
k} }}n_{\bf k} } , \label{eq:three}
\end{equation}
where  $n_{\bf k}$ is the number of particles in a state with a
momentum ${\bf k}$. At zero temperature we obtain:
\begin{equation}
\tilde t_\mathbf{p}  = t_\mathbf{p}  + c\frac{{V_0  + V_\mathbf{p}
}} {{1 + \delta _\mathbf{p}^0 }},
\label {eq:four}
\end{equation}
where $c = N/\Omega$ is the concentration of atoms in the system.
When we obtained (\ref{eq:four}), we assumed that at zero
temperature all the atoms are in a state with a momentum
$\mathbf{k}\text{ = 0}$, because this approximation takes into
account only the diagonal terms.

Obviously, $V_0\geqslant 0$: otherwise, the system would be
unstable \cite{1}. On the other hand, $V_{\bf q}$ is the regular
function of the parameter ${\bf q}$. So, we obtain the gap $\Delta
_0 = V_0 c$ in the spectrum of one-atom excitations of the system.

When Bogolyubov obtained this result, he thought that it was the
result of a wrong description of the system. Let us consider in
more detail the reason why Bogolyubov considered the appearance of
the gap in the spectrum of one-atom excitations a wrong solution.

Analysis of article  \cite{4} shows that Bogolyubov expected
$\mathop {\lim }\limits_{p \to 0} \tilde t_\mathbf{p} /p$ to
coincide with the sound speed in the system and thus have a finite
value. However, there is no physical principle, from which it
would follow that the rate of one-atom excitations and the speed
of the sound in the boson system should be equal, because the
phonons and one-atom excitations are two various physical
entities. To clarify this point it is sufficient to consider the
system above the transition point.

As for the other arguments of Bogolyubov, they are secondary and
required to explain why the obtained solution, from his point of
view, was wrong. These arguments come down to the fact that the
perturbation theory becomes inapplicable at small excitation
momenta. We think that this argument are unjustified, because it
is easy to show that the appearance of the gap in the one-atom
excitation spectrum makes the calculation of the spectrum on the
basis of the perturbation theory at small excitation momenta more
robust. Indeed, the diagonal part of Hamiltonian $\hat V$ , which,
in fact, is responsible for the appearance of the gap, can be
included in the so-called principle Hamiltonian of the
perturbation theory.  Accordingly, if this fact is taken into
account, the absolute value of the appearing energy denominators
turns out to be not less than the value $\sim \Delta _\mathbf{p} =
V_\mathbf{p} c$, which makes the mentioned calculation more
robust.

\section{\label{sec:level1}Phonons and One-atom Excitations in the Proposed Approach}

Let us consider an elastic medium, the quantum oscillations of
which correspond to creation and annihilation operators $\tilde
h_\mathbf{q}^ +$ and $\tilde h_\mathbf{q}$. This leads to a new
term $\hat H_{ph}$ to be added to the system Hamiltonian:
\begin{equation}
\hat H_{ph}  = \sum\limits_\mathbf{q} {sqh_\mathbf{q}^ +
h_\mathbf{q} } ,
\end{equation}
where $s$  -is the speed of the sound in the elastic medium thus
introduced. Let us name hereafter the quantum oscillations of the
elastic medium corresponding to the operators $\tilde
h_\mathbf{q}^ +$ and $\tilde h_\mathbf{q}$ seed phonons.

Let us introduce formally transformations corresponding to the
procedure of renormalization of the atom-phonon interaction:
\begin{equation}
\begin{gathered}
  b_\mathbf{p}^ +   = \tilde b_\mathbf{p}^ +   - \Omega ^{ - 1/2} \sum\limits_{\mathbf{q} \ne 0} {\varphi _\mathbf{q}^\mathbf{p} \tilde b_{\mathbf{p} - \mathbf{q}}^ +  \tilde h_\mathbf{q}^ +  }  + \Omega ^{ - 1/2} \sum\limits_{\mathbf{q} \ne 0} {\varphi _\mathbf{q}^{\mathbf{p} + \mathbf{q}*} \tilde b_{\mathbf{p} + \mathbf{q}}^ +  \tilde h_\mathbf{q} }  - \frac{1}
{2}\Omega ^{ - 1} \sum\limits_{\mathbf{q} \ne 0} {\left| {\varphi _\mathbf{q}^\mathbf{p} } \right|^2 } \tilde b_\mathbf{p}^ +   \hfill \\
   + \frac{1}
{2}\Omega ^{ - 1} \sum\limits_{\mathbf{k},\mathbf{q} \ne 0} {\left( {\varphi _{ - \mathbf{q}}^{\mathbf{p} - \mathbf{q}*} \varphi _{ - \mathbf{q}}^\mathbf{k}  - \varphi _\mathbf{q}^{\mathbf{k} + \mathbf{q}*} \varphi _\mathbf{q}^\mathbf{p} } \right)\tilde b_{\mathbf{p} - \mathbf{q}}^ +  \tilde b_{\mathbf{k} + \mathbf{q}}^ +  \tilde b_\mathbf{k} }  \hfill \\
   - \frac{1}
{2}\Omega ^{ - 1} \sum\limits_{\mathbf{k} \ne 0,\mathbf{q} \ne 0} {\left( {\varphi _\mathbf{q}^{\mathbf{p} - \mathbf{k} + \mathbf{q}*} \varphi _\mathbf{k}^\mathbf{p}  + \varphi _\mathbf{q}^{\mathbf{p} + \mathbf{q}*} \varphi _\mathbf{k}^{\mathbf{p} + \mathbf{q}} } \right)\tilde b_{\mathbf{p} + \mathbf{q} - \mathbf{k}}^ +  \tilde h_\mathbf{k}^ +  \tilde h_\mathbf{q} }  \hfill \\
   + \frac{1}
{4}\Omega ^{ - 1} \sum\limits_{\mathbf{k} \ne 0,\mathbf{q} \ne 0} {\left( {\varphi _\mathbf{q}^{\mathbf{p} - \mathbf{k}} \varphi _\mathbf{k}^\mathbf{p}  + \varphi _\mathbf{k}^{\mathbf{p} - \mathbf{q}} \varphi _\mathbf{q}^\mathbf{p} } \right)\tilde b_{\mathbf{p} - \mathbf{k} - \mathbf{q}}^ +  \tilde h_\mathbf{k}^ +  \tilde h_\mathbf{q}^ +  }  \hfill \\
   + \frac{1}
{4}\Omega ^{ - 1} \sum\limits_{\mathbf{k} \ne 0,\mathbf{q} \ne 0} {\left( {\varphi _\mathbf{q}^{\mathbf{p} + \mathbf{q}*} \varphi _\mathbf{k}^{\mathbf{p} + \mathbf{q} + \mathbf{k}*}  + \varphi _\mathbf{k}^{\mathbf{p} + \mathbf{k}*} \varphi _\mathbf{q}^{\mathbf{p} + \mathbf{q} + \mathbf{k}*} } \right)\tilde b_{\mathbf{p} + \mathbf{q} + \mathbf{k}}^ +  \tilde h_\mathbf{k} \tilde h_\mathbf{q} },  \hfill \\
\end{gathered}
\label {eq:six}
\end{equation}
\begin{equation}
\begin{gathered}
  h_\mathbf{q}^ +   = \tilde h_\mathbf{q}^ +   + \Omega ^{ - 1/2} \sum\limits_\mathbf{p} {\varphi _\mathbf{q}^{\mathbf{p}*} \tilde b_\mathbf{p}^ +  \tilde b_{\mathbf{p} - \mathbf{q}} }  \hfill \\
   + \frac{1}
{2}\Omega ^{ - 1} \sum\limits_{\mathbf{p},\mathbf{k} \ne 0} {\left( {\varphi _\mathbf{q}^{\mathbf{p}*} \varphi _\mathbf{k}^{\mathbf{p} - \mathbf{q} + \mathbf{k}}  - \varphi _\mathbf{q}^{\mathbf{p} + \mathbf{k}*} \varphi _\mathbf{k}^{\mathbf{p} + \mathbf{k}} } \right)\tilde h_\mathbf{k}^ +  \tilde b_\mathbf{p}^ +  \tilde b_{\mathbf{p} + \mathbf{k} - \mathbf{q}} }  \hfill \\
   + \frac{1}
{2}\Omega ^{ - 1} \sum\limits_{\mathbf{p},\mathbf{k} \ne 0} {\left( {\varphi _\mathbf{q}^{\mathbf{p} - \mathbf{k}*} \varphi _\mathbf{k}^{\mathbf{p}*}  - \varphi _\mathbf{k}^{\mathbf{p} - \mathbf{q}*} \varphi _\mathbf{q}^{\mathbf{p}*} } \right)\tilde h_\mathbf{k} \tilde b_\mathbf{p}^ +  \tilde b_{\mathbf{p} - \mathbf{k} - \mathbf{q}} } . \hfill \\
\end{gathered}
\label {eq:seven}
\end{equation}
Operators $\tilde b^ +$ , $\tilde b$, $\tilde h^ +$ and  $\tilde
h$ introduced in (\ref{eq:six}-\ref{eq:seven}) are consistent with
the standard commutation relations for bosons with a precision of
$\sim \varphi ^3$. Parameter $\varphi$ in this approach plays the
role of a formal small parameter.

In RPA approximation we formulate the equation on the
transformation parameters such a way that the system Hamiltonian
in terms of operators $\tilde b^ +$ , $\tilde b$, $\tilde h^ +$
and $\tilde h$ would not include terms of the structure $\tilde h^
+ \tilde b^ + \tilde b$ and $\tilde h\tilde b^ +  \tilde b$. This
equation will be of the form:
\begin{equation}
 - V_\mathbf{q} J_q  + \left( {t_{\mathbf{p} - \mathbf{q}}  - t_\mathbf{p}  + sq} \right)\varphi _\mathbf{q}^\mathbf{p}  + V_\mathbf{q} \left| {J_q } \right|^2 \varphi _\mathbf{q}^\mathbf{p}  + V_\mathbf{q} J_q^2 \varphi _{ - \mathbf{q}}^{\mathbf{p} - \mathbf{q}*}  = 0,
\label {eq:eight}
\end{equation}
where
\begin{equation}
J_q  = \Omega ^{ - 1} \sum\limits_\mathbf{p} {\left( {\varphi
_\mathbf{q}^{\mathbf{p} + \mathbf{q}}  - \varphi
_\mathbf{q}^\mathbf{p} } \right)n_\mathbf{p} } ,
\label {eq:nine}
\end{equation}
and $n_\mathbf{p}$ is the thermodynamic average of the number of
atoms in the state with a momentum $\mathbf{p}$. In (\ref{eq:eight})
the numerical coefficient with terms $\sim V_\mathbf{q} \left| {J_q
} \right|^2$, $\sim V_\mathbf{q} J_q^2$ was chosen in such a way
that the spectrum calculated in the procedure of renormalization
(see (\ref{eq:fifteen})) coincides with the spectrum of sound
obtained on the basis of analysis of the spectral dependence of the
system dielectric function \cite{5}. Actually, in this case, we
simply use the opportunity of choosing the order in which the terms
are considered in a situation where the system is missing a real
small parameter.

A nontrivial solution of equations (\ref{eq:six}-\ref{eq:seven})
under the assumption that parameter $\varphi _\mathbf{q}^\mathbf{p}$
is real-valued is given below:
\begin{equation}
\varphi _\mathbf{q}^\mathbf{p}  = \frac{{J_q V_\mathbf{q} \left(
{t_\mathbf{p}  - t_{\mathbf{p} - \mathbf{q}}  + sq} \right)}}
{{\omega _\mathbf{q}^2  - \left( {t_\mathbf{p}  - t_{\mathbf{p} -
\mathbf{q}} } \right)^2 }},
\end{equation}
\begin{equation}
J_q^2  = \frac{{\omega _\mathbf{q}^2  - s^2 q^2 }} {{2sqV_\mathbf{q}
}}.
\end{equation}
At that, $\omega _\mathbf{q}$ is consistent with the following
equation:
\begin{equation}
1 + \Omega ^{ - 1} \sum\limits_\mathbf{p} {V_\mathbf{q}
\frac{{n_\mathbf{p}  - n_{\mathbf{p} + \mathbf{q}} }}
{{t_{\mathbf{p} + \mathbf{q}}  - t_\mathbf{p}  + \omega _\mathbf{q}
}}}  = 0.
\label {eq:twelve}
\end{equation}
Thus, the dependence of $\omega _\mathbf{q}$ gives a sound spectrum
obtained by analyzing the dielectric function of the system.

Kinetic terms $\hat H_{ph}^{kin}$ contributing to the
renormalization of energy of the seed phonons calculated with the
use of RPA are of the form:
\begin{equation}
\hat H_{ph}^{kin}  = \sum\limits_\mathbf{q} {sq\tilde h_\mathbf{q}^
+  \tilde h_\mathbf{q} }  + \hat H_V  + \hat H_t ,
\end{equation}
where
\begin{equation}
\begin{gathered}
  \hat H_V  = \sum\limits_\mathbf{q} {V_\mathbf{q} J_q^2 \tilde h_\mathbf{q}^ +  \tilde h_\mathbf{q} }  + \frac{1}
{2}\sum\limits_\mathbf{q} {V_\mathbf{q} J_q^2 \tilde h_\mathbf{q}^ +
\tilde h_{ - \mathbf{q}}^ +  }  + \frac{1}
{2}\sum\limits_\mathbf{q} {V_\mathbf{q} J_q^2 \tilde h_\mathbf{q} \tilde h_{ - \mathbf{q}} } , \hfill \\
  \hat H_t  = \Omega ^{ - 1} \sum\limits_{\mathbf{p},\mathbf{q}} {\left( {\left( {t_{\mathbf{p} + \mathbf{q}}  - t_\mathbf{p}  - sq} \right)\left| {\varphi _\mathbf{q}^{\mathbf{p} + \mathbf{q}} } \right|^2  + \left( {t_{\mathbf{p} - \mathbf{q}}  - t_\mathbf{p}  + sq} \right)\left| {\varphi _\mathbf{q}^\mathbf{p} } \right|^2 } \right)n_\mathbf{p} \tilde h_\mathbf{q}^ +  \tilde h_\mathbf{q} }  \hfill \\
   + \Omega ^{ - 1} \sum\limits_{\mathbf{p},\mathbf{q}} {\left( {t_\mathbf{p}  - t_{\mathbf{p} + \mathbf{q}} } \right)\varphi _\mathbf{q}^{\mathbf{p} + \mathbf{q}} \varphi _{ - \mathbf{q}}^\mathbf{p} n_\mathbf{p} \tilde h_\mathbf{q}^ +  \tilde h_{ - \mathbf{q}}^ +  }  \hfill \\
   + \Omega ^{ - 1} \sum\limits_{\mathbf{p},\mathbf{q}} {\left( {t_\mathbf{p}  - t_{\mathbf{p} + \mathbf{q}} } \right)\varphi _\mathbf{q}^{\mathbf{p} + \mathbf{q}} \varphi _{ - \mathbf{q}}^\mathbf{p} n_\mathbf{p} \tilde h_\mathbf{q} \tilde h_{ - \mathbf{q}} } . \hfill \\
\end{gathered}
\end{equation}
For clarity we will first take into account only the contribution
of $\hat H_V$ to  $\hat H_{ph}^{kin}$, which, actually, as is
shown below, corresponds to the calculation of the spectrum in the
long-wave limit. Diagonalization of the operator $\hat
H_{ph}^{kin}$ in this case using the Bogolyubov transformation
leads to the following expression for the renormalized energy of
the seed phonons $\varepsilon _{\bf q}$:
\begin{equation}
\varepsilon _{\bf q}  = \omega _{\bf q} ,
\label {eq:fifteen}
\end{equation}
As is expected, the spectrum of the seed phonons coincides with the
expression obtained for the spectrum of sound vibrations by
calculating the dielectric function. In particular, for the Bose gas
at zero temperature $\omega _{\bf q}$ coincides with the well-known
expression obtained by Bogolyubov for the spectrum of quasiparticles
\cite{1} with neglect of the difference between the squared absolute
value of the so-called Bogolyubov $c$ - number and the full number
of atoms in the system, i.e. we have:
\begin{equation}
\varepsilon _{\bf q}  = \sqrt {t_{\bf q} \left( {t_{\bf q}  +
2\Delta _{\bf q} } \right)} .
\end{equation}
The resulting expressions for the renormalized quantities in the
suggested approach should be taken in the limit $s \to 0$. Indeed,
in this case $\hat H_{ph}$ does not add terms to the system
Hamiltonian. So, the system of seed phonons only gives a set of
states, which are used for the quantization of its oscillations. In
particular, for the additional term of interatomic interaction
$\delta \hat V$ following from the terms of Hamiltonian $\hat T$ and
$\hat H_{ph}$, we can obtain:
\begin{equation}
\delta \hat V = \frac{1} {2}\Omega ^{ - 1} \sum\limits_{{\bf p},{\bf
k},{\bf q}} {\delta V_{\bf q}^{{\bf p},{\bf k}} \tilde b_{{\bf p} +
{\bf q}}^ +  \tilde b_{{\bf k} - {\bf q}}^ +  \tilde b_{\bf k}
\tilde b_{\bf p} } ,
\end{equation}
where
\begin{equation}
\delta V_{\bf q}^{{\bf p},{\bf k}}  =  - \frac{1} {4}V_{\bf q}
\frac{{\omega _{\bf q}^2 \left[ {\left( {t_{{\bf p} + {\bf q}}  -
t_{\bf p} } \right)^2  + \left( {t_{\bf k}  - t_{{\bf k} - {\bf q}}
} \right)^2 } \right]}} {{\left( {\omega _{\bf q}^2  - \left(
{t_{{\bf p} + {\bf q}}  - t_{\bf p} } \right)^2 } \right)\left(
{\omega _{\bf q}^2  - \left( {t_{\bf k}  - t_{{\bf k} - {\bf q}} }
\right)^2 } \right)}}. \label
{eq:eighteen}
\end{equation}
Considering equation (\ref{eq:eighteen}) leads (in the limit
$p,k\rightarrow 0$) to the conclusion that renormalization does
not contribute to the size of the gap in the one-atom excitation
spectrum.

Now let us take into account also the contribution of $\hat H_t$
in the spectrum of the phonons for the case of zero temperature. A
simple calculation gives the following expression (compare with
(\ref{eq:fifteen})):
\begin{equation}
\varepsilon _{\bf q}  = \omega _{\bf q} \sqrt {1 + \frac{{t_{\bf q}
}} {{2\Delta _{\bf q} }}} .
\end{equation}
In fact, one can say that in the system one more expansion
parameter appears, which is equal to $t_{\bf q} /2\Delta _{\bf
q}$. It can be assumed, therefore, that the spectrum of phonons
ends at the values of wave vector ${\bf q}$ corresponding to the
condition $t_{\bf q} \sim 2\Delta _{\bf q}$. Recall in this
context that the spectrum of phonons is also limited at the
shortwave region when the Boltzmann distribution for the
occupation numbers of the atomic energy levels is determined.
Besides, at high temperatures $T > \Delta _{\bf q} /4$ equation
(\ref{eq:twelve}) has no solution for $\omega _{\bf q}$.

The suggested method makes it possible in principle to obtain the
properties of the system with any accuracy, if one does not take
into account the arbitrariness in the order of consideration of the
terms appearing in the renormalization procedure. This arbitrariness
is caused by the lack of a real small parameter in the system. We
plan to eliminate it by comparison with the results of alternative
methods of system's investigation.

Of course, not all the renormalized values will be calculated so
simply as the correction to interatomic interaction energy $\delta
\hat V$, because transformation parameter  $\varphi _{\bf q}^{\bf
p}$ has a singularity at  $s \to 0$. At the same time, it seems
that this problem is not of a fundamental nature. It is sufficient
to increase the accuracy of transformations
(\ref{eq:six}-\ref{eq:seven}). At that, the appearing singular
expressions will be reduced leaving only the finite values in the
calculation results.

In summary, one can assert that at this stage of development of
the suggested approach no reasons can be seen why the coexistence
of the two branches of the excitation spectrum of the system may
be doubted at temperatures below the transition point. It is quite
another matter that the results obtained in the context of the
suggested approach should be confirmed by other methods. Of
course, in this case we mean a description of the system with the
use of the technique of Green functions modified as compared with
\cite{6}. This technique should not break the fundamental
symmetries of the system. We plan to perform such consideration in
near future.

\section{\label{sec:level1}Bose Gas Thermodynamics of One-atom Excitations with Interaction}

Now let us consider the case of finite temperatures assuming that
they are sufficiently low to carry out the condition
\begin{equation}
\sum\limits_{\mathbf{k} \ne 0} {V_{\mathbf{k} - \mathbf{p}}
n_\mathbf{k} }  \approx V_\mathbf{p} N_e, \label{eq:twenty}
\end{equation}
where  $N_e$ is the full number of atoms in the excited states. In
this case the simplification of (\ref{eq:three}) gives the
following expressions:
\begin{equation}
\begin{gathered}
  \tilde t_0  = 2\Delta _0  - \kappa \Delta _0,  \hfill \\
  \tilde t_\mathbf{p}  = t_\mathbf{p}  + 2\Delta _0  + \left( {\Delta _\mathbf{p}  - \Delta _0 } \right)\left\{ {\text{at }\mathbf{p} \ne 0} \right\}, \hfill \\
\end{gathered}
\end{equation}
where  $\kappa$ is the fraction of atoms condensed in the ground
orbital.

Let us assume that in the case of thermally activated atoms it is
possible to use the parabolic dispersion law with a mass $M_e$,
that is
\begin{equation}
\tilde t_{\bf p}  \approx 2\Delta _0  + \frac{{p^2 }} {{2M_e }}
\end{equation}
at ${\bf p} \ne 0$. This simplification is possible, because $V_{\bf
p}  - V_0  \sim p^2$ at ${\bf p} \to 0$.

As for the system energy $E$ , using the expression for $\hat
H_{diag}$ gives:
\begin{equation}
E = \sum\limits_{\bf p} {t_{\bf p} n_{\bf p} }  + \frac{{V_0 }}
{{2\Omega }}\sum\limits_{\bf p} {n_{\bf p} \left( {n_{\bf p}  - 1}
\right)}  + \frac{1} {2}\Omega ^{ - 1} \sum\limits_{{\bf p} \ne {\bf
k}} {\left( {V_0  + V_{{\bf k} - {\bf p}} } \right)n_{\bf p} n_{\bf
k} }
\end{equation}

Again, assuming that the temperatures are sufficiently low to
satisfy the conditions
\begin{equation}
\sum\limits_{{\bf k} \ne 0,{\bf p} \ne 0,{\bf p} \ne {\bf k}}
{V_{{\bf k} - {\bf p}} n_{\bf p} n_{\bf k} }  \approx V_0
\sum\limits_{{\bf k} \ne 0,{\bf p} \ne 0,{\bf p} \ne {\bf k}}
{n_{\bf p} n_{\bf k} }  \approx V_0 N_e^2 , \label{eq:twentyfour}
\end{equation}
it is possible to calculate energy $\varepsilon$ per one atom of
the large system:
\begin{equation}
\varepsilon  = \Delta _0 \left( {1 - \frac{{\kappa ^2 }} {2}}
\right) + \left( {1 - \kappa } \right)\left\langle t \right\rangle
_e ,
\label {eq:twentyfive}
\end{equation}
where $\left\langle t \right\rangle _e$ is the average kinetic
energy of the excited atoms of the system:
\begin{equation}
\left\langle t \right\rangle _e  = N_e^{ - 1} \sum\limits_{{\bf p}
\ne 0} {t_{\bf p} \left\langle {n_{\bf p} } \right\rangle },
\end{equation}
For average occupation numbers $\left\langle {n_\mathbf{p} }
\right\rangle$ the Bose-Einstein distribution function is
applicable. Thus, we have:
\begin{equation}
\left\langle {n_\mathbf{p} } \right\rangle  = \frac{1} {{\exp \left(
{\left( {p^2 /2M_e  + \kappa \Delta _0 } \right)/T} \right) - 1}},
\end{equation}
because we assume, as usual, that the value of the chemical
potential of the system is very close to $\tilde t_0$.

Pass from sums to integrals it is possible to obtain an equation
for the fraction of condensed atoms:
\begin{equation}
\kappa  = 1 - \frac{{\left( {2M_e T} \right)^{3/2} }} {{4\pi ^2 }c}
\int {\frac{{\sqrt x dx}} {{\exp \left( {\kappa \Delta_0 /T + x}
\right) - 1}}}
\label {eq:twentyeight}
\end{equation}
and an expression for $\left\langle t \right\rangle _e$:
\begin{equation}
\left\langle t \right\rangle _e  = \frac{{M_e }} {M}T\frac{{\int
{\frac{{x\sqrt x dx}} {{\exp \left( {\kappa \Delta _0 /T + x}
\right) - 1}}} }} {{\int {\frac{{\sqrt x dx}} {{\exp \left( {\kappa
\Delta _0 /T + x} \right) - 1}}} }}.
\end{equation}
Accordingly, it becomes possible to compare experimental data with
calculations based on the proposed scheme and to estimate the gap
$\Delta_0$ and mass $M_e$ of the excited atoms of the system.

\section{\label{sec:level1}Determination of Model Parameters Based on Experimental Data
with the Use of Enthalpy and the Fraction of the SF Component in He
II}

The calculation idea consists in the possibility of using two
equations ~(\ref{eq:twentyfive}, \ref{eq:twentyeight}) in order to
determine two parameters $\Delta_0$ and  $\gamma  = M_e /M$ at known
values of temperature $T$, energy  $\varepsilon$, and fraction of
atoms condensed at the ground orbital $\kappa$.

The value of energy $\varepsilon$ per one atom of helium can be
obtained from the data of Kapitsa on the temperature dependence of
enthalpy \cite{7}, because under the conditions of an isobaric
experiment it is very close in value to the temperature dependence
of the internal energy of the system. Indeed, pressure in the
experiments of Kapitsa does not exceed 25 atm. So, the difference
between the enthalpy of helium at zero and finite temperatures is
very close in value to the difference of the corresponding
internal energy values, because density changes in He II caused by
temperature are very small.

According to the proposed concept the total enthalpy of the system
determined in the experiment is a sum of two contributions: phonon
- roton and one-atom excitations. Now we have a task of extraction
the contribution of one-atom excitations from the total energy of
the system.

The phonon contribution to the internal energy of the system is
connected with the temperature dependence of the phonon heat
capacity by the relation:
\begin{equation}
U_{ph} \left( T \right) = \int\limits_0^T {C_{ph} \left( t
\right)dt}.
\end{equation}

\begin{table}
\caption{\label{tab:table2}Basic data and calculation results for a
temperature range  $1.5 - 2.1 K$}
\begin{ruledtabular}
\begin{tabular}{lccccccc}
 $T, K$&$1.5$&$1.6$&$1.7$&$1.8$&$1.9$&$2.0$&$2.1$\\
\hline $U_{tot}$, cal/g \cite{7}& 0.071 & 0.112 & 0.176 &0.252& 0.351 & 0.484 & 0.662 \\
$U_{ph}$, cal/g \cite{8}& 0.046 & 0.068 & 0.095 &0.129& 0.170 & 0.219 & 0.276 \\
$\kappa \text{, \% }$  \cite{9} & 94.33 & 89.83 & 83.92 &76.12& 66.90 & 55.79 & 41.84 \\
$U_{at}$, cal/g& 0.025 & 0.044 & 0.081 &0.123& 0.181 & 0.265 & 0.386 \\
$\tilde \Delta$, $K$& 0.19 & 0.08 & 0.04 &0.01& - & - & - \\
$\tilde \gamma$& 0.43 & 0.52 & 0.63 &0.70& - & - & - \\
\end{tabular}
\end{ruledtabular}
\end{table}

At that, there is the well-known expression \cite{8} for the heat
capacity  $C_{ph} \left( t \right)$ :
\begin{equation}
\begin{gathered}
  C_{ph} \left( t \right) = \frac{{2\pi ^2 }}
{{15}}k_B \left( {\frac{{k_B t}}
{{\hbar S}}} \right)^3  \hfill \\
   + k_B \frac{{p_0^2 \sqrt {\mu _r k_B t/2} }}
{{\pi ^{3/2} \hbar ^3 }}\left( {\frac{{\Delta _r^2 }} {{k_B^2 t^2 }}
+ \frac{{\Delta _r }} {{k_B t}} + \frac{3} {4}} \right)\exp \left( {
- \frac{{\Delta _r }}
{{k_B t}}} \right), \hfill \\
\end{gathered}
\label {eq:thirtyone}
\end{equation}
where $k_B$ - is Boltzmann constant. Nowadays experimental values of
the phonon spectrum parameters used in (\ref{eq:thirtyone}) are as
follows:
\begin{equation}
\begin{gathered}
  S = 2.38 \cdot 10^4 \text{ cm/sec}\text{,} \hfill \\
  \Delta _r  = 8.6\text{ K}\text{,} \hfill \\
  p_0  = 1.8 \cdot 10^8 \text{ }\hbar /\text{cm}, \hfill \\
  \mu _r  = 0.13\text{ }M. \hfill \\
\end{gathered}
\end{equation}
Obviously, within the limits of a simple model of two branches of
excitations one-atom excitation energy $U_{at} \left( T \right)$ can
be obtained by simple subtraction of the phonon-roton contribution
from the total energy of the system $U_{tot} \left( T \right)$:
\begin{equation}
U_{at} \left( T \right) = U_{tot} \left( T \right) - U_{ph} \left( T
\right).
\label {eq:thirtythree}
\end{equation}
Temperature dependence of the fraction of condensed atoms in the
ground state $\kappa \left( T \right)$ can be directly obtained
from the results of an experiment by Andronikashvili (measuring
the period of axial-torsion oscillations of a vessel filled with
He II \cite{9}, or that by Peshkov (measuring the speed of second
sound \cite{10}). Both experiments give very similar dependencies
$\kappa \left( T \right)$. It is relevant to note that the
fundamentals of the phenomenological theory of He II, including
the possibility of a two-fluid description of the system, cannot
be doubted.

Table 1 presents the basic data for calculation of  $U_{at} \left( T
\right)$ , as well as the results of calculations of other
parameters that will be determined below. Fig. 1, respectively,
presents the dependence of $\delta \varepsilon \left( T \right)$,
value equal to energy $U_{at} \left( T \right)$ calculated per one
atom of helium. Note here that the energy $\delta \varepsilon$,
obviously, should be equal to energy $\varepsilon$, counted from its
own value at $T = 0\text{ }K$, i.e. (compare with
~(\ref{eq:twentyfive}))
\begin{equation}
\delta \varepsilon  = \Delta _0 \left( {1 - \kappa ^2 } \right)/2 +
\left( {1 - \kappa } \right)\gamma T\frac{{\int {\frac{{x\sqrt x
dx}} {{\exp \left( {\kappa \Delta _0 /T + x} \right) - 1}}} }}
{{\int {\frac{{\sqrt x dx}} {{\exp \left( {\kappa \Delta _0 /T + x}
\right) - 1}}} }}.
\label {eq:thirtyfour}
\end{equation}

\begin{figure}
\includegraphics{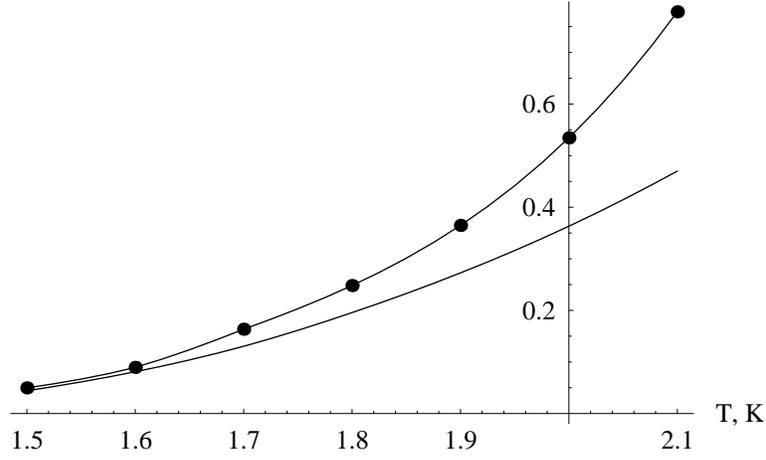}
\caption{\label{fig:epsart}Temperature dependence of one-atom
excitation energy $\delta \varepsilon$ calculated from atomic energy
$U_{at}$  per one helium atom (upper curve). Besides, a calculated
curve for the dependence $\delta \varepsilon \left( T \right)$ on
the basis of~(\ref{eq:thirtyfour}) at selected values of $\Delta_0$
and $\gamma$ determined by~(\ref{eq:thirtysix}) is presented (lower
curve). All data are given in Kelvins.}
\end{figure}

Dependencies $\tilde \Delta \left( T \right)$ and  $\tilde \gamma
\left( T \right)$ presented in the table are determined with the use
of a self-consistent system of two equations (compare with
~(\ref{eq:twentyfive}, \ref{eq:twentyeight}))
\begin{equation}
\begin{gathered}
  \delta \varepsilon \left( T \right) = \tilde \Delta \left( T \right)\left( {1 - \kappa ^2 \left( T \right)} \right)/2 \hfill \\
   + \left( {1 - \kappa \left( T \right)} \right)\tilde \gamma \left( T \right)T\frac{{\int {\frac{{x\sqrt x dx}}
{{\exp \left( {\kappa \left( T \right)\tilde \Delta \left( T
\right)/T + x} \right) - 1}}} }} {{\int {\frac{{\sqrt x dx}}
{{\exp \left( {\kappa \left( T \right)\tilde \Delta \left( T \right)/T + x} \right) - 1}}} }}, \hfill \\
  \kappa \left( T \right) = 1 - \frac{{\left( {2\tilde \gamma \left( T \right)MT} \right)^{3/2} }}
{{4\pi ^2 c}} \int {\frac{{\sqrt x dx}}
{{\exp \left( {\kappa \left( T \right)\tilde \Delta \left( T \right)/T + x} \right) - 1}}} . \hfill \\
\end{gathered}
\end{equation}

The obtained dependencies $\tilde \Delta \left( T \right)$ and
$\tilde \gamma \left( T \right)$ are the basis for obtaining the
parameters $\Delta_0$ and  $\gamma$ of the model. Indeed, when we
deduced formula (\ref{eq:twentyfive}) we made the
assumptions~(\ref{eq:twenty}, \ref{eq:twentyfour})), which are
realized the better the lower is the system temperature. On the
other hand, it is difficult for us to estimate the temperature
starting at which assumptions (\ref{eq:twenty},
\ref{eq:twentyfour}) are realized with acceptable accuracy,
because there are no reliable data on interatomic potentials. In
this situation in order to obtain parameters $\Delta_0$ and
$\gamma$ it is necessary to extrapolate dependencies $\tilde
\Delta \left( T \right)$ and $\tilde \gamma \left( T \right)$ to
the zero temperature range, where these assumptions are realized
perfectly.

\begin{figure}
\includegraphics{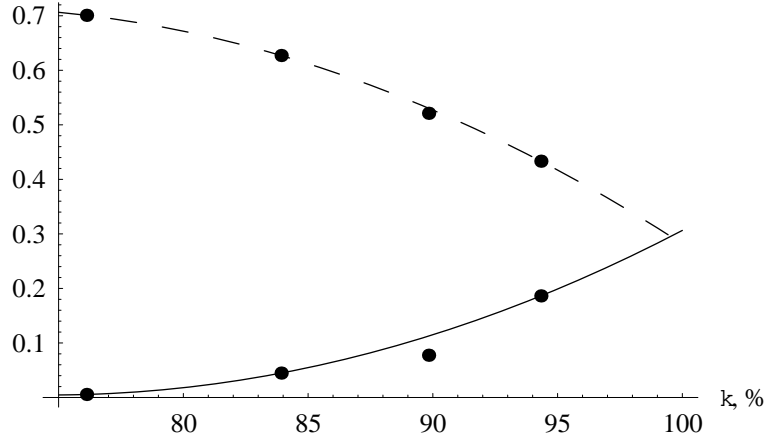}
\caption{\label{fig:epsart}Calculated dependencies $\tilde \Delta
\left( \kappa  \right)$ (solid curve, $\tilde \Delta$ are given in
Kelvins) and $\tilde \gamma \left( \kappa  \right)$ (dashed curve).}
\end{figure}
On the other hand, using the dependency $\kappa \left( T \right)$
enables to obtain the dependencies  $\tilde \Delta \left( \kappa
\right)$ and $\tilde \gamma \left( \kappa  \right)$ shown in Fig.
2. At that, it becomes possible to extrapolate the obtained
functions to the range of $\kappa  = 1$. This also corresponds to
the zero temperature case. This option seems to be more
attractive, because the extrapolation becomes non-dimensional, and
its value of $\approx \text{5\% }$ seems to be small.
Extrapolation parameters $\Delta_0$ and  $\gamma$ obtained in this
case have the following values:
\begin{equation}
\begin{gathered}
  \Delta _0  = 0.31\text{ K}\text{,} \hfill \\
  \gamma  = 0.28. \hfill \\
\end{gathered}
\label {eq:thirtysix}
\end{equation}
It is obvious that dependency $\delta \varepsilon \left( T \right)$,
calculated on the basis of~(\ref{eq:thirtyfour}) at the values of
parameters $\Delta_0$ and  $\gamma$ selected in~(\ref{eq:thirtysix})
should approach dependency $\delta \varepsilon \left( T \right)$
calculated on the basis of ~(\ref{eq:thirtythree}) at decreasing
temperature, because in this case correlations (\ref{eq:twenty},
\ref{eq:twentyfour}) are realized better. Fig. 1 visualizes this
statement.

\section{\label{sec:level1}Calculation of the Dielectric Function of the System}

When calculating the system response to an external action one
should keep in mind the requirement of gradient invariance of the
model. The below calculation takes into account this circumstance
by using the well-known fact used for calculating the Hartree
dielectric function: if Hamiltonian $\hat T$ is taken as the basic
Hamiltonian of the system, it is possible to calculate the
response on the external potential with use of the standard theory
of excitations. The Hamiltonian of this external potential $\hat
U$ has the form
\begin{equation}
\hat U = \int {U\left( {{\bf r},t} \right)b^ +  \left( {\bf r}
\right)b\left( {\bf r} \right)} ,
\label {eq:thirtyseven}
\end{equation}
where  $b^ +  \left( {\bf r} \right)$ and  $b\left( {\bf r} \right)$
are creation an annihilation operators in coordinate representation.
In this case the gradient symmetry of the system is not broken.

As one goes to a new basis of wave functions, in order to solve
this problem in the context of the perturbation theory, a number
of terms appear that have the same structure in RPA as the right
hand side of equation (\ref{eq:thirtyseven}). This circumstance is
taken into account by introducing effective potential $U_{eff}
\left( {{\bf r},t} \right)$. At that, for Fourier transforms
$U\left( {{\bf r},t} \right)$ and $U_{eff} \left( {{\bf r},t}
\right)$ the following correlation exists:
\begin{equation}
U_{eff} \left( {{\bf q},\omega } \right) = \frac{{U\left( {{\bf
q},\omega } \right)}} {{\varepsilon _H \left( {{\bf q},\omega }
\right)}},
\end{equation}
where  $\varepsilon _H \left( {{\bf q},\omega } \right)$ is the
Hartree dielectric function:
\begin{equation}
\varepsilon _H \left( {{\bf q},\omega } \right) = 1 + V_{\bf q}
\sum\limits_{\bf k} {\frac{{n_{\bf k}  - n_{{\bf k} + {\bf q}} }}
{{t_{{\bf k} + {\bf q}}  - t_{\bf k}  - \left( {\omega  + i0}
\right)}}} .
\label {eq:thirtynine}
\end{equation}
The presented procedure does not take into account the residual
Hamiltonian $\hat H_{res}$, for which we have:
\begin{equation}
\hat H_{res}  = \hat V + \hat U - \hat U_{eff} ,
\end{equation}
where
\begin{equation}
\hat U_{eff}  = \int {U_{eff} \left( {{\bf r},t} \right)b^ +  \left(
{\bf r} \right)b\left( {\bf r} \right)d{\bf r}} .
\end{equation}
In the new basis of wave functions, obtained after solving the
Hartree problem of the perturbation theory, the matrix elements
$\hat H_{res}$ are invariant under the gradient transformation.
Further let us restrict consideration to a simple case of the
diagonal terms.

This consideration leads to a shift of the system energy levels.
It turns to be important for determination of the occupation
numbers $n_{\bf k}$ and $n_{{\bf k} + {\bf q}}$ in
(\ref{eq:thirtynine}): calculating them one should use the
Bose-Einstein distribution function, in which one-atom energies
$\tilde t_{\bf k}$ and $\tilde t_{{\bf k} + {\bf q}}$ are used as
energy levels.

Note one more feature of the Hamiltonian $\hat H_{res}$. Obviously,
the response of the system on the Hamiltonian $\hat U_{eff}$
described by the dielectric function at zero temperature corresponds
to the response of an ideal SF fluid. At the same time,
non-diagonal, time-dependent terms of the Hamiltonian $\hat H_{res}$
can lead to dissipation of energy due to transitions from the ground
state to the excited ones if the frequency of the external action
exceeds the gap energy in the one-atom excitation spectrum. Here it
is reasonable to make a most general remark: the appearance of
time-dependent non-diagonal terms of the Hamiltonian in the
procedure of calculating the system response to an external action
for the models using gapless one-atom excitation spectra can mean
the appearance of an energy dissipation channel in the system, even
if it satisfies the SF principles from a formal point of view.

On the other hand, the hypothetical possibility of the existence
of a gap in the one-atom excitation spectrum of SF state has been
suggested before \cite{11,12}. For such models it is possible to
observe directly the gap size values. For example, if voltage is
applied to a metal needle, then near its tip a force appears that
exerts influence upon the helium atoms. Therefore, the conditions
of the experiment proposed earlier in very general terms
\cite{12}, become better defined, because now we know the
frequency interval, in which a thorough investigation of the phase
and amplitude-frequency characteristics of the chain which
includes a sensor in the form of the above needle should be
carried out.
\section{\label{sec:level1}Conclusion}

This paper suggests the SF model of He II preserving the
fundamental symmetries of the system at all temperatures including
those below the transition point. The structure of energy levels
appearing in the model is determined by the possibility of
coexistence of phonon and one-atom spectra below the transition
point.

Numerical simulation showed satisfactory agreement of the model
with the experimental data below the transition point. In
particular, we succeeded in estimating the value of the gap in the
one-atom excitation spectrum, which turned out to be approximately
equal to 0.31 K at zero temperature. We suggest a method of direct
observation of the gap in the one-atom excitation spectrum on
basis of the data on energy absorption in the system under the
influence of an external oscillating force.

\begin{acknowledgments}
The authors express their gratitude to A.A.Rukhadze for his
unfailing attention towards the problem and a moral support.
\end{acknowledgments}

\bibliography{Superfluidity}

\end{document}